\documentclass[aoas,preprint]{imsart}

\RequirePackage[OT1]{fontenc}
\RequirePackage{amsthm,amsmath}
\RequirePackage{natbib}
\RequirePackage[colorlinks,citecolor=blue,urlcolor=blue]{hyperref}

\arxiv{1710.04560}
\usepackage{multirow}
\usepackage{amssymb}
\usepackage{graphicx}
\usepackage{graphicx}
\usepackage{comment}
\usepackage{mathrsfs}
\usepackage[colorinlistoftodos]{todonotes}
\usepackage{rotating}
\usepackage{lscape}
\usepackage{epstopdf}
\usepackage{natbib}
\newtheorem{theorem}{Theorem}
\newtheorem{corollary}{Corollary}

\newcommand{\ind}{\stackrel{\mathrm{ind}}{\sim}}
\newcommand{\N}{\mathrm{N}}
\newcommand{\E}{\mathrm{E}}
\usepackage{lipsum}
\def\undertilde#1{\mathord{\vtop{\ialign{##\crcr
$\hfil\displaystyle{#1}\hfil$\crcr\noalign{\kern1.5pt\nointerlineskip}
$\hfil\tilde{}\hfil$\crcr\noalign{\kern1.5pt}}}}}

\begin{document}
\begin{frontmatter}
\title{Bayesian Modeling of the Structural Connectome for Studying Alzheimer's Disease}

\begin{aug}
\author{\fnms{Arkaprava} \snm{Roy}\thanksref{t1, m4}\ead[label=e1]{arkaprava.roy@duke.edu}},
\author{\fnms{Subhashis} \snm{Ghosal}\thanksref{t1, m1}\ead[label=e2]{sghoshal@ncsu.edu}},
\author{\fnms{Jeffrey} \snm{Prescott}\thanksref{m2}
\ead[label=e3]{jwprescott@gmail.com}}
\author{\fnms{Kingshuk} \snm{Roy Choudhury}\thanksref{m4}
\ead[label=e4]{kingshuk.roy.choudhury@duke.edu}}
\and
\author{\fnms{for the Alzheimer's Disease Neuroimaging Initiative}\thanksref{t2}}

\thankstext{t1}{Research is partially supported by NSF grant DMS-1510238.}
\thankstext{t2}{Data used in the preparation of this article were obtained from the Alzheimer's Disease Neuroimaging Initiative
(ADNI) database (adni.loni.usc.edu). As such, the investigators within the ADNI contributed to the design
and implementation of ADNI and/or provided data but did not participate in analysis or writing of this report.
A complete listing of ADNI investigators can be found at:
\url{http://adni.loni.usc.edu/wp-content/uploads/how_to_apply/ADNI_Acknowledgement_List.pdf}}
\runauthor{Roy et al.}
\runtitle{Bayesian Modeling for Studying Alzheimer's Disease}

\affiliation{Duke University\thanksmark{m4}, North Carolina State University\thanksmark{m1} and
MetroHealth Medical Center\thanksmark{m2}}

\address{Arkaprava Roy\\
Department of statistics\\
Duke University\\
Durham, NC\\
Subhashis Ghosal\\
Department of statistics\\
North Carolina State University\\
Raleigh, NC\\
\printead{e1}\\
\phantom{E-mail:\ }\printead*{e2}}

\address{Jeffrey Prescott\\
Dept. of Radiology\\
MetroHealth Medical Center\\
Cleveland, OH\\
Kingshuk Roy Choudhury\\
Dept. of Biostatistics and Bioinformatics\\
Duke University Medical Center\\
Durham, NC\\
\printead{e3}\\
\phantom{E-mail:\ }\printead*{e4}}
\end{aug}

\begin{abstract}
We study possible relations between Alzheimer's disease progression and the structure of the connectome which is white matter connecting different regions of the brain. Regression models in covariates including age, gender, and disease status for the extent of white matter connecting each pair of regions of the brain are proposed. Subject inhomogeneity is also incorporated in the model through random effects with an unknown distribution. As there is a large number of pairs of regions, we also adopt a dimension reduction technique through graphon (\cite{Lovasz}) functions, which reduces the functions of pairs of regions to functions of regions. The connecting graphon functions are considered unknown but the assumed smoothness allows putting priors of low complexity on these functions. We pursue a nonparametric Bayesian approach by assigning a Dirichlet process scale mixture of zero to mean normal prior on the distributions of the random effects and finite random series of tensor products of B-splines priors on the underlying graphon functions. We develop efficient Markov chain Monte Carlo techniques for drawing samples for the posterior distributions using Hamiltonian Monte Carlo (HMC). The proposed Bayesian method overwhelmingly outperforms a competing method based on ANCOVA models in the simulation setup. The proposed Bayesian approach is applied on a dataset of 100 subjects and 83 brain regions and key regions implicated in the changing connectome are identified.
\end{abstract}


\end{frontmatter}

\section{Introduction}
\label{introduction}

Alzheimer's disease (AD) is a neurodegenerative disorder that affects approximately 5 million people in the US and 30 million. Current thought is that detecting pathologic changes in the brain before the development of clinical symptoms will allow for a successful treatment. In particular, one area of active investigation is how changes in white matter, which constitutes the brain’s structural network or ‘connectome’, occur alongside Alzheimer’s disease progression \citep{phillips2015graph}. Recent research has found that changes in structural connectivity become more widespread with Alzheimer’s disease progression \citep{tucholka2018structural}  and that changes in the connectome appear to preferentially affect less connected areas of the brain \citep{daianu2015rich}. Further, changes in white matter are correlated with amyloid plaque burden, which is one of the pathological hallmarks of the AD and has been shown to become elevated years before the onset of clinical symptoms \citep{Prescott}. From a methodological viewpoint, all these papers rely on modelling specific scalar metrics, such as connectivity indices, which are derived from the connectivity graphs obtained via image processing brain scans (described later in methods). One drawback of this approach is that the conclusions can depend on the choice of metric \citep{phillips2015graph}. In this paper, we overcome this dilemma by modelling changes to the entire graph of the network.

Our study is performed using data obtained by Alzheimer's Disease Neuroimaging Initiative (ADNI $-$ \url{adni.loni.usc.edu}). Using T1-weighted magnetic resonance (MR) images from ADNI, the cortex of the brain is divided into several regions using a standard anatomic atlas. These regions are connected by white matter fibers, identified using diffusion tensor imaging (DTI) MR. The data used in this study were from the ADNI 2 study and included all subjects with diffusion-weighted imaging as of November 2012. Readers can be referred to the Materials and Methods section of \cite{Prescott} for details of MR image acquisition and processing. The graphical representation of these white matter connections between cortical regions is referred to as the connectome. It is thought that some of these connections between brain regions become weakened over time due to the AD. Some other factors like age or sex might also affect the connectome, as well as subject-specific random effects. We model the connectome using graph-theoretic metrics, accounting for patients' specific effects and implement a Bayesian analysis. 

Here, we consider a connectome with 83 cortical regions in the brain. Mathematically, the connectome can be viewed as a graph $(V, E)$, where $V$ denotes the set of nodes or vertices standing for brain regions and $E$ for the edges between pairs of vertices whenever present. As in a graphical model, edges are marked with certain measurements. In our context, the measurements consist of observing the presence or absence of white matter fibers connecting two regions, the number of white matter fiber and the mean length of white matter fibers between them. Our aim is to identify the significant pairs of regions corresponding to different covariates in the connectome. The aspects of the presence of a connection, the number of connections and the mean length are modeled respectively by a binary regression model with a probit link, a Poisson regression model with an exponential link and a normal regression model. Interactions of covariates with pairs of regions are considered. Because the number of region-pairs is prohibitively high, it leads to a very high dimensional regression model if a naive approach is taken. Hence we use a dimension reduction technique that introduces a latent variable for each region and expresses functions of region-pairs in terms of a single smooth unknown function of each pair of latent variables as in a graphon model (\cite{Lovasz}). If $(a_{jk}: j,k=1,\ldots,J)$ is an array of parameters, then we model  $a_{jk}=g(u_j,u_k)$, where $u_1,\ldots,u_J$ are latent variables and $g$ is a function of two arguments. Symmetry of the matrix $(\!(a_{jk})\!)$ is respected if $g$ is symmetric in its arguments. The original motivation for this representation is that if $a_{jk}$, $j,k=1,\ldots,J$, are random variables such that the matrix $(\!(a_{jk})\!)$ is distributionally invariant under permutations of rows and columns, then $a_{jk}=g(u_j,u_k)$ for some latent variables $u_1,\ldots,u_J$ that can be assumed to be uniformly distributed and for a function $g$, to be called a graphon function. In the present context, if the parameters $a_{jk}$, $j,k=1,\ldots,J$, are treated as random, then such row and column wise exchangeability conditions are natural non-informativeness conditions. Given the graphon function, thus the strength of connections is determined by only 83 latent variables linked with each region instead of being $\binom{83}{2}=3486$ making huge computational savings. The graphon function is also treated as an unknown without any specific parametric form and is nonparametrically estimated from the data. More specifically, we put a finite random series prior based on tensor products of B-splines, where the coefficients are given appropriate prior distributions. Finite random series priors are widely used in the literature to construct priors on various functions and are systematically studied, for instance, in \cite{Shen}, but it seems to have not been used before for putting prior distributions on graphon functions. The assumed smoothness of the graphon function helps keep the number of basis function required for the basis expansion relatively small. This is because to approximate a function in $[0,1]^d$ of smoothness index $\mathfrak{f}$ by a finite series within accuracy $\epsilon$, one needs to use only $O(\epsilon^{-d/\mathfrak{f}})$ many elements of standard bases like polynomials, B-splines or wavelets, that is fewer functions are needed for smoother functions. The proposed Bayesian procedure can be shown to lead to consistent posterior, in the setting where the number of regions remains fixed but the number of subjects increases to infinity. The result goes beyond the particular application we are addressing in the present paper. Although the proof uses the standard hypothesis testing and prior positivity of Kullback-Leibler neighborhoods approach developed by Schwartz, our major technical contribution is the construction of exponentially consistent tests for random effects in a Poisson regression model, along with weakening conditions on the predictor variables to include a larger variety of applications. Details on posterior consistency are shown in the supplementary material. 

The rest of the paper is organized as follows. In the next section, we describe the details of modeling the connectome dataset from ADNI. In Section~\ref{prior}, we describe the prior construction and develop posterior computing techniques. A simulation study comparing the proposed Bayesian procedure with ANCOVA-based ones is conducted in Section~\ref{simulation}. The real-data on connectome from ADNI is analyzed in Section~\ref{real-data}. Finally Section~\ref{conclusions-discussion} ends with some concluding remarks.

\section{Data description and modeling}
\label{data description}

Data used in the preparation of this article were obtained from the Alzheimer’s Disease Neuroimaging Initiative (ADNI) database (\url{adni.loni.usc.edu}). The ADNI was launched in 2003 as a public-private partnership, led by Principal Investigator Michael W. Weiner, MD. The primary goal of ADNI has been to test whether serial magnetic resonance imaging (MRI), positron emission tomography (PET), other
biological markers, and clinical and neuropsychological assessment can be combined to measure the progression of mild cognitive impairment (MCI) and early Alzheimer’s disease (AD). Here, we examine if the connectome can be used as a biomarker. In the ADNI dataset on connectome, the brain is divided into $J=83$ regions. For each pair of brain regions, the number of white matter fibers between them and their average lengths is obtained. 
The data is obtained for $n=100$ subjects, for whom information regarding disease prognosis, sex and age are also available. For many edges, the mean lengths are not defined where there are no white matter fibers. There are three disease prognosis states, Alzheimer (AD), mild cognitive impairment (MCI) and no cognitive impairment (NC). 

\begin{figure}[htbp]
\centering
\includegraphics[width = 1\textwidth]{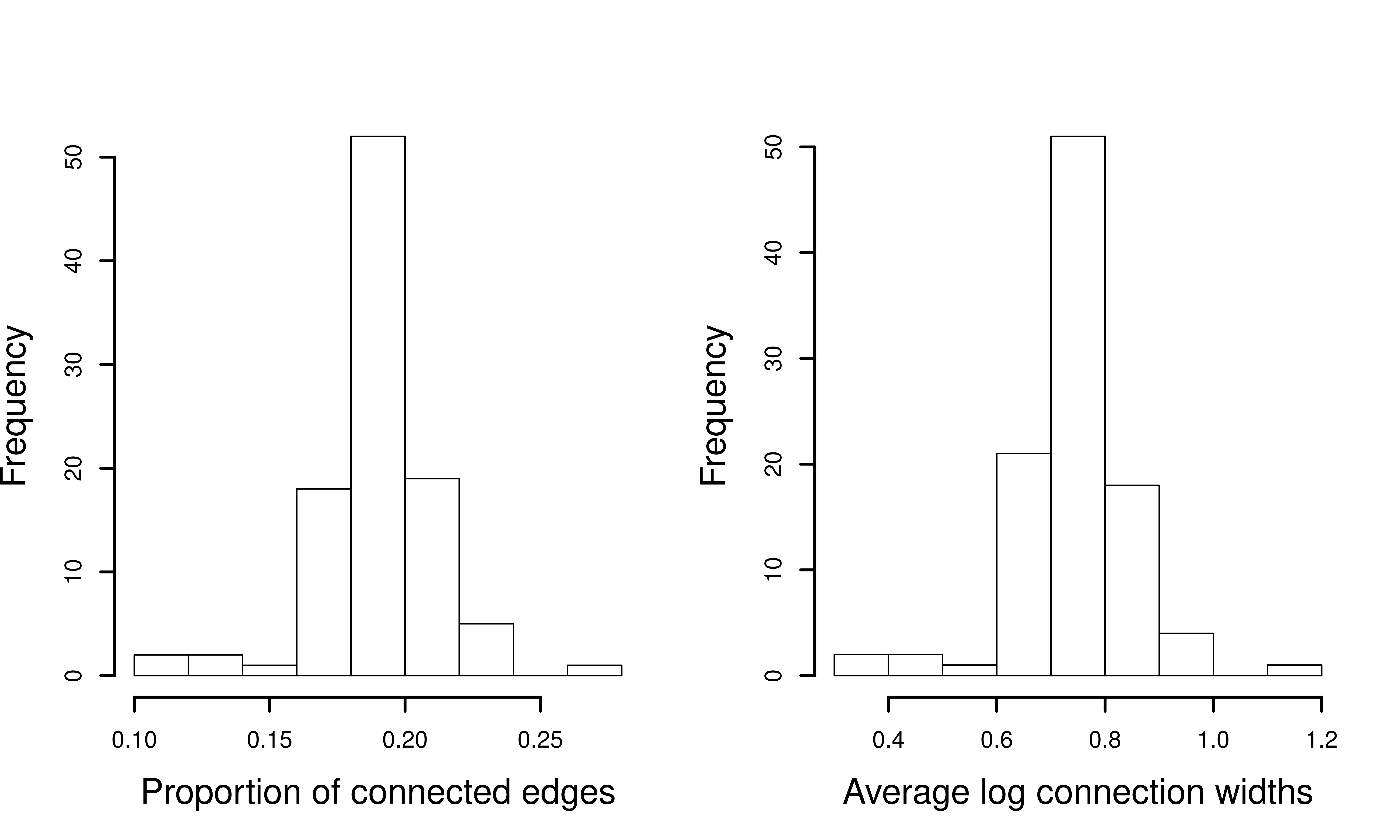}
\caption{Histogram of proportion of connected edges and average log connection widths across individual. {\tt dip.test} of R package {\tt diptest} suggests that these are unimodal with kurtosis ($\mu_4/\mu_2^2 - 3$) of 3.74 and 3.79 respectively in comparison to kurtosis of normal, being zero.}
\label{heatmap}
\end{figure}

Except for age, the other two covariates, sex and disease prognosis, are categorical. Since the disease prognosis has three possible states, we introduce dummy variables $Z^{\mathrm{MCI}}$ and $Z^{\mathrm{AD}}$ respectively standing for the onset of MCI and AD, setting NC at the baseline. Similarly, the dummy variable $Z^{\mathrm{M}}$ indicating male gender is introduced setting females at the baseline. Let $Z=(Z^{\mathrm{MCI}},Z^{\mathrm{AD}},Z^{\mathrm{M}},\mathrm{Age})'$ stand for the whole vector of covariates and $Z_i$ stand for its value for the $i$th subject. Let $N_{ijk}$ stand for the number of white matter fiber connecting brain regions $j$ and $k$ in the $i$th subject, and $L_{ijk}$ the mean length of such fibers, provided that $N_{ijk}\ge 1$. In ADNI dataset, there were no missing values in $N_{ijk}$  or  $L_{ijk}$. For three individuals gender and age were missing. They were imputed using R package {\tt MICE}.

It seems natural to consider a Poisson model for the counts of fibers connecting two regions in a subject. However, as shown in Figure~\ref{heatmap}, the proportions of connected edges among individuals are between 10\% to 30\%. Thus, the abundance of zero connections makes the Poisson model somewhat inappropriate. We overcome the problem by considering a zero-inflated Poisson model, by boosting the probability of zero through a binary latent variable $\Xi_{ijk}$ with parameter $\Phi(\pi_{ijk})$, where $\Phi$ stands for the standard normal distribution function and $\pi_{ijk}$ is a real-valued parameter. If $\Xi_{ijk}=0$, then  $N_{ijk}$ is set at zero, while if $\Xi_{ijk}=1$, the number of connections $N_{ijk}$ is assumed to be Poisson distributed with some positive mean $e^{\lambda_{ijk}}$. Note that in our formulation $\Xi_{ijk}$ is not completely identifiable since the value $N_{ijk}=0$ is compatible with both possible values of $\Xi_{ijk}$. If $N_{ijk}\ge 1$, we assume that the mean fiber length, in the logarithmic scale, is normally distributed with some mean $\mu_{ijk}$, and variance $\sigma^2/N_{ijk}$ for some unknown $\sigma>0$. The heuristic justification of the choice of the variance $\sigma^2/N_{ijk}$ stems from the fact that $L_{ijk}$ is an average of $N_{ijk}$ (independent) variables and should be approximately normal with variance inversely proportional to the number $N_{ijk}$ of averaging variables. Since fiber lengths are positive, the model seems to fit the data better in the logarithmic scale, and the heuristics for the choice of the variance extends to the logarithmic scale by the delta method, at least when $N_{ijk}$ is large. Thus we can represent the data generating process as 
\begin{gather}
\Xi_{ijk} \sim \mathrm{Bin}(1,\Phi(\pi_{ijk}))\nonumber\\ 
N_{ijk}|\{\Xi_{ijk}=1\} \sim \mathrm{Poisson}(e^{\lambda_{ijk}}) \label{basic model}\\
\log L_{ijk} = \mu_{ijk} + \epsilon_{ijk}, \qquad \epsilon_{ijk}|\{N_{ijk}\ge 1\} \sim \mathrm{N}(0, \sigma^2/N_{ijk}).\nonumber
\end{gather}

A simple analysis of covariance (ANCOVA)-type model can be formulated to describe linear effects of the covariates $Z_i$ on each unrestricted parameter $\pi_{ijk}$, $\lambda_{ijk}$ and $\mu_{ijk}$ for each pair of brain regions $(j,k)$:
\begin{gather}
\mu_{ijk} = (\!(\mu_{0})\!)_{j,k} + Z_i'\chi_{jk} + \eta_{1i},\nonumber\\
\pi_{ijk} = (\!(\pi_{0})\!)_{j,k} + Z_i'\beta_{jk} + \eta_{2i},\label{regression}\\
\lambda_{ijk} = (\!(\lambda_{0})\!)_{j,k} +  Z_i'\nu_{jk} + \eta_{3i},\nonumber
\end{gather}
where $(\!(\mu_{0})\!)_{j,k}$, $(\!(\pi_{0})\!)_{j,k}$ and $(\!(\lambda_{0})\!)_{j,k}$ are baseline values of $\mu_{ijk}$, $\pi_{ijk}$ and $\lambda_{ijk}$ respectively for covariate value $Z_i=0$. Let
\begin{gather}
\chi_{jk}=((\!(\chi_{\mathrm{MCI}})\!)_{j,k},(\!(\chi_{\mathrm{AD}})\!)_{j,k},(\!(\chi_{\mathrm{M}})\!)_{j,k},(\!(\chi_{\mathrm{Age}})\!)_{j,k})',\nonumber\\ \beta_{jk}=((\!(\beta_{\mathrm{MCI}})\!)_{j,k},(\!(\beta_{\mathrm{AD}})\!)_{j,k},(\!(\beta_{\mathrm{M}})\!)_{j,k},(\!(\beta_{\mathrm{Age}})\!)_{j,k})', \label{array of coefficients}\\ \nu_{jk}=((\!(\nu_{\mathrm{MCI}})\!)_{j,k},(\!(\nu_{\mathrm{AD}})\!)_{j,k},(\!(\nu_{\mathrm{M}})\!)_{j,k},(\!(\nu_{\mathrm{Age}})\!)_{j,k})'\nonumber 
\end{gather}
be regression coefficients for the average length, connection probability and number of connections respectively and $(\!(.)\!)_{j,k}$ denote $(j,k)^{th}$ element of a matrix,   
and $\eta_i=(\eta_{1i}, \eta_{2i}, \eta_{3i})'$, $i=1,\ldots,n$, be independent random effects of the $i$th subject distributed according to an unknown common distribution. It may be noted that the normal distribution function $\Phi$ and the exponential  function are used respectively as link functions for binary and Poisson regression. For the latter, the exponential link is almost a universal choice, while for binary regression both logistic and probit (i.e. $\Phi$) links are commonly used and usually give similar results. Our preference for the probit link is due to its computational advantage in a Gibbs sampling scheme for Bayesian computation, through a data-augmentation technique (see \cite{Albert}).

For a preliminary analysis, we fit the model using a generalized heteroscedastic ANCOVA, ignoring the zero-inflation aspect and the random effects in the model. The model thus has $3\times \binom{83}{2}=10458$ parameters and $34860$ observations of mean length and number of white matter fibers corresponding to $100$ subjects and $3486$ potential edges between different brain regions. We observed that for several edges $(j,k)$, the maximum likelihood method failed to give estimates of either $\mu_{0,jk}$ or $\chi_{jk}$. For the Poisson regression, the {\tt glm} function in R did not converge for several pairs $(j,k)$. This is due to an insufficient number of observations. Thus it suggests using a dimension reduction of the parameter space through further modeling if we want to conduct an edge-wise analysis. The dimension reduction also helps with computation and gives easy interpretability of the results. 

Since the parameters are indexed by edges, a substantial reduction of dimension will be possible if these can be viewed as arising from some latent characteristics of nodes through some fixed but unknown function. This can be motivated from exchangeability considerations. In the absence of initial information about connections between regions, exchangeability seems to be an appealing assumption. By a well known representation theorem of exchangeable random graphs (c.f. \cite{Aldous}, \cite{Hoover}), a function of edge $(j,k)$ can then be represented as $f(\xi_i,\xi_j)$ where $\xi_i$, for each node $i$ is a latent variable independently and identically distributed and $f$ is a fixed function, called a graphon, irrespective of the size of the network. Assuming that the function $f$ is sufficiently smooth, a basis expansion can approximate it using only a few terms. Thus the graphon technique in our context will be able to reduce a parameter array of size $\binom{83}{2}=3486$ to only a parameter vector of size $83+K$, where $K$ is the number of parameters used to approximate the unknown smooth graphon function. Typically a modest number of terms suffices for well-behaved functions using standard bases such as B-splines or polynomials. As a result, a substantial dimension reduction is possible through the graphon technique. This leads to modeling the arrays of baseline values and regression coefficient as 
\begin{gather}
(\!(\mu_{0})\!)_{j,k} = \mu(\xi_j,\xi_k), \nonumber\quad 
(\!(\pi_{0})\!)_{j,k} = \pi(\xi_j, \xi_k), \nonumber\quad 
(\!(\lambda_{0})\!)_{j,k} = \lambda(\xi_j, \xi_k),\\
(\!(\chi_l)\!)_{j,k} = \chi_{l}(\delta_{j}, \delta_{k}), \quad l=\text{MCI, AD, M, Age},\label{graphon}\\
(\!(\beta_{l})\!)_{j,k} = \beta_{l}(\delta_{j}, \delta_{k}), \quad l=\text{MCI, AD, M, Age},\nonumber\\
(\!(\nu_{l})\!)_{j,k} = \nu_{l}(\delta_{j}, \delta_{k}), \quad l=\text{MCI, AD, M, Age},\nonumber
\end{gather}
where, with an abuse of notations, $\mu$, $\pi$, $\lambda$, $\chi_{\mathrm{MCI}}$, $\chi_{\mathrm{AD}}$, $\chi_{\mathrm{M}}$, $\chi_{\mathrm{Age}}$,  $\beta_{\mathrm{MCI}}$, $\beta_{\mathrm{AD}}$, $\beta_{\mathrm{M}}$, $\beta_{\mathrm{Age}}$,  $\lambda_{\mathrm{MCI}}$, $\lambda_{\mathrm{AD}}$, $\lambda_{\mathrm{M}}$ and $\lambda_{\mathrm{Age}}$ are smooth functions on the unit square $[0,1]^2$ and symmetric in their arguments, and $\xi_1,\ldots,\xi_J$ and $\delta_1,\ldots,\delta_J$ are latent variables taking values in the unit interval. The reason for choosing two separate sets of latent variables $\xi$ and $\delta$ is to distinguish between fixed and main effects.


\section{Prior specification and posterior computation}
\label{prior}

\subsection{Prior specification}

To proceed with a nonparametric Bayesian analysis, we put prior distributions on the smooth functions appearing in the graphon representation through basis expansion in tensor products of B-splines, and on the coefficients of the basis expansion. The coefficients can be arranged in the form of a square matrix. The symmetry of the matrices of coefficients ensures symmetry of the resulting functions in its arguments as required by graphon functions. Given other sets of parameters and values of the random effects, (independent) normal priors on the coefficients of the tensor products of B-splines will lead to conjugacy in the normal regression model for the length, allowing a simple and fast posterior updating rule. In the binary regression model for the connection probability, normal prior still leads to conjugacy using the data augmentation technique of \cite{Albert}. Since no conjugacy is possible for the Poisson regression for the number of connections, gradient-based Hamiltonian Monte Carlo sampling algorithm is applied. Alternatively, adaptive rejection sampling can be applied to obtain posterior updates. On the distribution $G$ of the random effects, we put a Dirichlet process scale mixture of zero mean normal prior (see Chapter 5 of \cite{Ghosal} and \cite{west1987scale} for scale mixture of normal). The histogram plot of Figure~\ref{heatmap} as well as Figure 1 from the supplementary materials motivate us that the distributions of random effects are symmetric but non-normal with higher kurtosis.

More specifically, the prior can be completely described by the following set of relations: 
\begin{itemize}
\item [(i)] Graphon functions: 
\begin{gather*}
\mu(\xi_j, \xi_k) = \sum_{m=1}^K \sum_{m'=1}^K \theta_{1,mm'} B_m(\xi_j)B_{m'}(\xi_k),\\
\pi(\xi_j, \xi_k) =  \sum_{m=1}^K \sum_{m'=1}^K\theta_{2,mm'} B_m(\xi_j)B_{m'}(\xi_k),\\
\lambda(\xi_j, \xi_k) = \sum_{m=1}^K \sum_{m'=1}^K \theta_{3,mm'} B_m(\xi_j)B_{m'}(\xi_k),
\end{gather*}
and for $l=\mathrm{MCI, AD, M}$, 
\begin{gather*}
\chi_{l}(\delta_{j}, \delta_{k}) = \sum_{m=1}^K \sum_{m'=1}^K \gamma_{1l,mm'}B_m(\delta_j)B_{m'}(\delta_k),\\
\beta_{l}(\delta_{j}, \delta_{k}) = \sum_{m=1}^K \sum_{m'=1}^K \gamma_{2l,mm'}B_m(\delta_j)B_{m'}(\delta_k),\\
\nu_{l}(\delta_{j}, \delta_{k}) = \sum_{m=1}^K \sum_{m'=1}^K \gamma_{3l,mm'}B_m(\delta_j)B_{m'}(\delta_k), 
\end{gather*}
where $\theta_{t,mm'} = \theta_{t,m'm}$ for all $t = 1,2,3,$ and $\gamma_{tl,mm'} = \gamma_{t,m'm}$ for all $t = 1,2,3$, $l=\mathrm{MCI, AD, M}$, and that 
\begin{itemize}
\item [(a)] graphon coefficients: 
For some chosen $a>0$, $$\theta_{t,mm'} \ind \N (0, a^2), \quad \gamma_{tl,mm'} \ind \N (0, a^2),\quad 1\le m\le m'\le K,$$
for $t = 1,2,3$, $l=\mathrm{MCI, AD, M}$;
\item [(b)] latent variables:  $$\textrm{logit} (\xi_1),\ldots, \textrm{logit}(\xi_{J})\ind \N (0, a^2),\quad \textrm{logit}(\delta_1), \ldots, \textrm{logit}(\delta_{J})\ind \N (0, a^2),$$
here logit stands for the $\textrm{logit}(x)=\log(x/(1-x))$.  
\end{itemize}
\item [(ii)] Random effects distribution: For $t =  1,2,3$ and $i = 1,\ldots, n$, 
$$\eta_{ti}|\tau_{ti} \ind \mathrm{N}(0, \tau_{ti}^2), \quad 
\tau_{ti}^2\ind G_t, t=1,2,3, \quad G_t \ind \textrm{DP}(\alpha_t\textrm{IG}(b_1,b_2)),$$ 
where DP stands for the Dirichlet process, IG for the inverse-gamma distribution and the precision parameter $\alpha_t$ of the Dirichlet process is given a gamma prior 
$\alpha_t \sim \text{Ga}(c_1,c_2)$. 

\item [(iii)] Error variance: $\sigma^{-2}\sim \mathrm{Ga}(d_1,d_2)$.
\end{itemize}

\subsection{Posterior updating}

Introduce a latent variable $I_j$ the indicator of the Un(0,1) component of the distribution of $\delta_j$, $j=1,\ldots,J$. 
Now the conditional log-likelihood is given by 
\begin{align}
 \nonumber C& -\sum_{i,j,k} \exp\{\sum_{m,m'} [\theta_{3,mm'} B_m(\xi_j)B_{m'}(\xi_k)+ \sum_{l}\gamma_{3l,mm'} B_m(\delta_j) B_{m'}(\delta_k) Z_{il}] +  \nonumber\\ &\eta_{3i}\}+ \sum_{i,j,k}  N_{ijk} [\sum_{m,m} (\theta_{3,mm'} B_m(\xi_j)B_{m'}(\xi_k)+ \sum_{l}\gamma_{3l,mm'} B_m(\delta_j) B_{m'}(\delta_k) Z_{il})  \nonumber\\ &+ \eta_{3i}]+ 
- \frac{1}{2\sigma^2} \sum_{i,j,k} N_{ijk}[\log L_{ijk} -\sum_{m=1}^K \sum_{m'=1}^K (\theta_{1,mm'} B_m(\xi_j)B_{m'}(\xi_k) \nonumber\\&\qquad \qquad +\sum_{l}\gamma_{1l,mm'} B_m(\delta_j) B_{m'}(\delta_k) Z_{il}) + \eta_{1i}]^2
 \nonumber\\  &+ \sum_{i,j,k} I(N_{ijk} = 0) \log \Phi\big(\sum_{m,m'} (\theta_{2,mm'} B_m(\xi_j)B_{m'}(\xi_k)\nonumber\\&+ \sum_{l}\gamma_{2l,mm'} B_m(\delta_j) B_{m'}(\delta_k) Z_{il})+ \eta_{2i}\big) \nonumber\\ &+ \sum_{i,j,k}(1 - I(N_{ijk} = 0))\log\big(1 - \Phi\big(\sum_{m,m'} (\theta_{2,mm'} B_m(\xi_j)B_{m'}(\xi_k)\nonumber\\&+ \sum_{l}\gamma_{2l,mm'} B_m(\delta_j) B_{m'}(\delta_k) Z_{il})+ \eta_{2i}\big)\big) \nonumber\\  &- \frac{1}{2a^2}\sum_{m\le m'} (\theta_{1,mm'}^2 + \theta_{2,mm'}^2 + \theta_{3,mm'}^2) \nonumber-\frac{1}{2a^2} \sum_{m\leq m'}\sum_l (\gamma_{1,mm'l}^2 + \gamma_{2,mm'l}^2 + \gamma_{3,mm'l}^2) \nonumber\\  &+ \log\big((1 - I_j){\delta_j^{M-1}(1-\delta_j)^{M-1}} \frac{\Gamma(M)^2}{\Gamma(2M)}+ I_j\big)  \nonumber\\ 
&+ I_j\log q+ (1-I_j)\log (1-q) -(n J^2/2 + d_1-1) \log \sigma^2 - d_2/\sigma^2,\nonumber\\\label{logl}
\end{align}
where $C$ involes only hyperparameters $a, M, K, b_1, b_2, c_1, c_2, d_1, d_2,q$ and the observations,  but not the parameters of the model. 

The parameters having conjugacy are updated using Gibbs sampling scheme and most of the other parameters are updated using gradient based Hamiltonian Monte Carlo (HMC). Detailed calculations of all the posterior updates are shown in the supplementary material.  

\subsection{Tuning}
The leapfrog parameter is kept fixed at 10. The step length is tuned after each 100 iteration to ensure an acceptance rate between $55\%$ to $90\%$. The number of B-spline basis functions ($K$) is tuned via a grid search over a sequence of values in the range 7--20. For each possible values of $K$, we generate 10 sets of latent variables. For each set of latent variables, we can fit a simple linear regression to estimates the B-spine coefficients and calculate average the Akaike Information Criterion (AIC) over all the sets of latent variables. Based on these AIC values, we pick the $K$ with the lowest AIC value or the smallest value after which there is not much improvement in the AIC.

\section{Simulation}
\label{simulation}

In this section, we study the performance of the proposed Bayesian method in comparison with ANCOVA. For computational simplicity, we do not consider the random effects in the data generating process as well as in the model, that is, we consider the following analog of \eqref{basic model}:
\begin{gather}
\Xi_{ijk} \sim \mathrm{Bin}(1,\Phi(\pi_{ijk}))\nonumber\\ 
N_{ijk}|\{\Xi_{ijk}=1\} \sim \mathrm{Poisson}(e^{\lambda_{ijk}}) \nonumber\\
\log L_{ijk} = \mu_{ijk} + \epsilon_{ijk}, \qquad \epsilon_{ijk}|\{N_{ijk}\ge 1\} \sim \mathrm{N}(0, \sigma^2/N_{ijk}),\label{ancova}\\
\pi_{ijk} = (\!(\pi_{0})\!)_{j,k} + Z_i'\beta_{jk}\quad \mu_{ijk}=\mu_{0,jk}+Z_i'\chi_{jk}, \quad \lambda_{ijk}=\lambda_{0,jk}+Z_i'\nu_{jk}. \nonumber
\end{gather}

We consider $n=50, 100, 200, 500, 1000, 2000$ subjects for data generation with $J=20$ and $n=100, 250, 500, 1000$ for $J=40$ nodes. There is another simulation setup with sample size 100 and $J=80$ which is similar to our real-data application.

{\it{Data generation:}} 

The true matrices are generated as follows with $\epsilon_{l,jk}=(e_{l,jk}+e_{l,kj})/2$ and $e_{jk}\sim$N$(0, (\sqrt{0.05})^2)$ for $l=1,\ldots,15$,
\begin{align*}
\mu_0(j, k) &= ((-0.5\xi_j-0.4\xi_k)^3+(-0.4\xi_j-0.5\xi_k)^3)/2+\epsilon_{1,jk},\\
\pi_0(j,k) &= ((-0.7\xi_j-\xi_k)^3+(-\xi_j-0.7\xi_k)^3)/2+\epsilon_{2,jk}
,\\
\lambda_0(j,k) &= ((-0.5\xi_j-0.4\xi_k)^3+(-0.4\xi_j-0.5\xi_k)^3)/2+\epsilon_{3,jk},\\
\chi_\text{MCI}(j, k) &= (\exp(-0.5\delta_j-0.4\delta_k)+\exp(-0.4\delta_j-0.5\delta_k))/2+\epsilon_{4,jk},\\
\beta_\text{MCI}(j,k) &= (\exp(-0.7\delta_j-\delta_k)+\exp(-\delta_j-0.7\delta_k))/2+\epsilon_{5,jk},
\\
\nu_\text{MCI}(j,k) &= (\exp(-0.5\delta_j-0.4\delta_k)+\exp(-0.4\delta_j-0.5\delta_k))/2+\epsilon_{6,jk},\\
\chi_\text{AD}(j, k) &= (\sin(-0.5\delta_j-0.4\delta_k)+\sin(-0.4\delta_j-0.5\delta_k))/2+\epsilon_{7,jk},\\
\beta_\text{AD}(j,k) &= (\sin(-0.7\delta_j-\delta_k)+\sin(-\delta_j-0.7\delta_k))/2+\epsilon_{8,jk},
\\
\nu_\text{AD}(j,k) &= (\sin(-0.5\delta_j-0.4\delta_k)+\sin(-0.4\delta_j-0.5\delta_k))/2+\epsilon_{9,jk},\\
\chi_\text{M}(j, k) &= (\cos(-0.5\delta_j-0.4\delta_k)+\cos(-0.4\delta_j-0.5\delta_k))/2+\epsilon_{10,jk},\\
\beta_\text{M}(j,k) &= (\cos(-0.7\delta_j-\delta_k)+\cos(-\delta_j-0.7\delta_k))/2+\epsilon_{11,jk},
\\
\nu_\text{M}(j,k) &= (\cos(-0.5\delta_j-0.4\delta_k)+\cos(-0.4\delta_j-0.5\delta_k))/2+\epsilon_{12,jk},\\
\chi_\text{Age}(j, k) &= ((-0.5\delta_j-0.4\delta_k)+(-0.4\delta_j-0.5\delta_k))/2+\epsilon_{13,jk},\\
\beta_\text{Age}(j,k) &= ((-0.7\delta_j-\delta_k)+(-\delta_j-0.7\delta_k))/2+\epsilon_{14,jk},
\\
\nu_\text{Age}(j,k) &= ((-0.5\delta_j-0.4\delta_k)+(-0.4\delta_j-0.5\delta_k))/2+\epsilon_{15,jk}.
\end{align*}

If $\Xi_{ijk}$ is generated as zero, the edge $(j,k)$ of $i^{th}$ individual will be missing. The generated data based on these functions have similar missingness compared to the real-data. We add the error component along with the functional values to deviate it a little bit from an exact functional form.
We have performed 50 replications for each case. We collect 5000 MCMC samples after burning in 5000 initial samples to draw the inference. The result for $J=40$ are based on 30 replications and 3000 post-burn samples after burning in 3000 initial samples.

{\it{Choice of the hyperparameters:}}
We choose the hyperparameters $a= 10$, $M=10$, $b_1=b_2=0.1$ and $c_1=c_2=10$ in \eqref{logl}. We take 7 B-spline basis functions based on the AIC values over a grid of possible number of B-spline basis functions. For all the simulated datasets, it always produces the optimal number as 7 as the number of nodes is fixed at 20.

For the ANCOVA based estimation, we use the weighted least squares technique for the normal model and the generalized linear regression for a Poisson regression model with the exponential link function. 

\begin{figure}[htbp]
\centering
\includegraphics[width=1\linewidth]{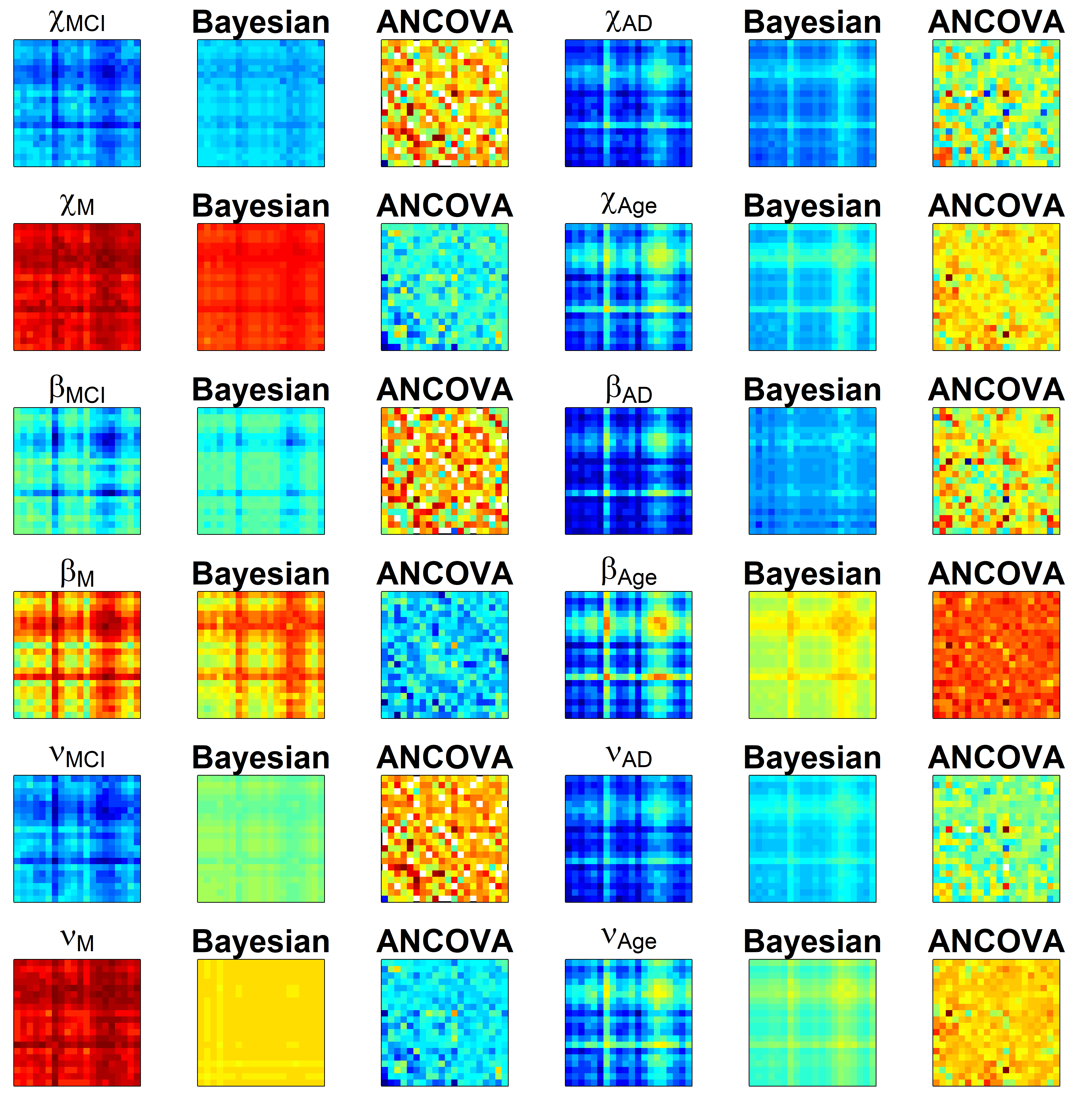}
\caption{Heatmap of the truth along with estimates from the Bayesian methods and the ANCOVA for sample size $2000$ with 20 nodes. First and fourth columns are the true parameter and subsequent columns present the corresponding Bayesian and ANCOVA estimates, averaged over 50 replications.}
\label{heatmapws}
\end{figure}

\begin{table}[htbp]
\caption{Comparison of estimation accuracy of order $10^{-2}$ for $n=2000$ with $J=20$}
\footnotesize\setlength{\tabcolsep}{1pt}
\centering
\begin{tabular}{|l|lll|lll|lll|}
  \hline
 & \multicolumn{3}{c|}{Bayes Graphon} & \multicolumn{3}{c|}{ANCOVA}& \multicolumn{3}{c|}{Bayes ANCOVA}\\
 & Square bias & Variance &MSE& Square bias & Variance &MSE& Square bias & Variance &MSE\\
   \hline
$\chi_\text{MCI}$ & 1.363  & 0.018  & 1.381  & 45.5  & 4.1  & 49.6 &52.181& 5.905& 58.087\\ 
  $\chi_\text{AD}$ & 1.057  & 0.011  & 1.068  & 25  & 2.63  & 27.63 &27.669& 3.845 & 31.515\\ 
  $\chi_\text{M}$ & 1.768  & 0.006  & 1.775  & 82.5  & 2.4  & 84.9 &82.162& 3.520 & 85.682\\ 
  $\chi_\text{Age}$ & 1.298  & 0.004  & 1.302  & 24.3  & 0.6  & 24.9 &25.806& 0.932 &26.739\\ 
  $\beta_\text{MCI}$ & 1.041  & 0.019  & 1.059  & 25.5  & 1.2  & 26.7 &40.037& 5.321& 45.358\\ 
  $\beta_\text{AD}$ & 1.825  & 0.012  & 1.837  & 61.3  & 1  & 62.3 & 77.604 & 4.069&81.674\\ 
  $\beta_\text{M}$ & 1.418  & 0.013  & 1.431  & 45.2  & 0.6  & 45.8 & 54.186 & 3.214 & 57.4\\ 
  $\beta_\text{Age}$ & 2.375  & 0.004  & 2.378  & 84.3  & 0.2  & 84.5 & 87.355 & 0.898 &88.254\\ 
  $\nu_\text{MCI}$ & 7.427  & 0.023  & 7.449  & 43.7  & 0.7  & 44.4 & 45.838 & 0.530 &46.368\\ 
  $\nu_\text{AD}$ & 3.549  & 0.015  & 3.563  & 23.2  & 0.5  & 23.7 & 24.249 & 0.366 & 24.615\\ 
  $\nu_\text{M}$ & 9.066  & 0.007  & 9.072  & 81.3  & 0.4  & 81.7 & 83.559 & 0.369 & 83.928\\ 
  $\nu_\text{Age}$ & 3.126  & 0.005  & 3.132  & 23.9  & 0.1  & 24  & 24.082 & 0.106 &24.188\\ 
   \hline
\end{tabular}
\label{tablews}
\end{table}

\begin{table}[htbp]
\caption{Comparison of estimation accuracy of order $10^{-2}$ for $n=200$ with $J=80$}
\footnotesize\setlength{\tabcolsep}{1pt}
\centering
\begin{tabular}{|l|lll|lll|}
  \hline
 & \multicolumn{3}{c|}{Bayes Graphon} & \multicolumn{3}{c|}{ANCOVA}\\
 & Square bias & Variance &MSE& Square bias & Variance &MSE\\
   \hline
$\chi_\text{MCI}$ & 2.58 & 0.04  & 2.63  & 74.16 & 94.93  & 169.09 \\ 
  $\chi_\text{AD}$ & 3.86 & 0.02  & 3.89  & 35.19 & 70.73  & 105.92 \\ 
  $\chi_\text{M}$ & 1.82 & 0.01  & 1.83  & 83.43 & 82.76  & 166.19 \\ 
  $\chi_\text{Age}$ & 4.16 & 0.01  & 4.18  & 30.10 & 19.07  & 49.17 \\ 
  $\beta_\text{MCI}$ & 2.63 &$3.35\times 10^{-02}$  & 2.63  & 16.66 & 0.92  & 17.58 \\ 
  $\beta_\text{AD}$ & 8.73 &$1.87\times 10^{-02}$  & 6.73 & 38.66 & 0.66  & 39.32 \\ 
  $\beta_\text{M}$ & 5.71 &$1.98\times 10^{-02}$  & 5.71  & 25.95 & 0.65  & 26.60\\ 
  $\beta_\text{Age}$ & 18.05 & $6.96\times 10^{-03}$  & 18.05  & 61.84 & 0.12  & 61.96 \\ 
  $\nu_\text{MCI}$ & 14.26 &$ 1.65\times 10^{-06} $ & 14.26  & 104.99 & 82.21  & 187.19 \\ 
  $\nu_\text{AD}$ & 8.36& $1.09\times 10^{-06}$  & 8.36  & 40.52 & 53.97  & 94.49 \\ 
  $\nu_\text{M}$ & 19.05 & $4.48\times 10^{-07} $ & 19.05  & 71.58 & 64.94 & 136.52 \\ 
  $\nu_\text{Age}$ & 8.86 & $5.17\times 10^{-07} $ & 8.86  &37.90 & 15.94  & 53.84 \\ 
   \hline
\end{tabular}
\label{tablews80}
\end{table}

\begin{table}[htbp]
\caption{Comparing prediction performances for different sample sizes across different models for J=20. Prediction MSE is used for mean-width and Predictive mean log-likelihood is used for the number of connections. Half of the recorded sample size is used to estimate model parameters and the rest half is used to evaluate prediction performance.}
\footnotesize\setlength{\tabcolsep}{1pt}
\centering
\resizebox{0.9\textwidth}{!}{\begin{minipage}{\textwidth}
\begin{tabular}{|l|l|l|l|l|l|l|l|l|l|}
  \hline
   & \multicolumn{3}{c|}{Bayes Graphon} & \multicolumn{3}{c|}{ANCOVA}& \multicolumn{3}{c|}{Bayes ANCOVA}\\\hline
 Total&&&&&&&&&\\ 
sample size& 2000 & 1000 &500& 2000 & 1000 &500& 2000 & 1000 &500\\\hline
 
 Mean-width &1.30&1.32&1.32&1.61&1.63&1.70&1.61&1.64&1.73\\ \hline Number of&$-168.61$&$-169.21$&$-210.26$&$-735.28$&$-835.21$&$-811.41$&$-181.43$&$-184.18$&$-288.50$\\ connections &&&&&&&&&\\
 
   \hline

   \hline
\end{tabular}
\end{minipage}}
\label{tablepred20}
\end{table}

\begin{table}[htbp]
\caption{Comparing prediction performances for different sample sizes across different models for J=40. Prediction MSE is used for mean-width and Predictive mean log-likelihood is used for the number of connections. Half of the recorded sample size is used to estimate model parameters and the rest half is used to evaluate prediction performance.}
\footnotesize\setlength{\tabcolsep}{1pt}
\centering
\resizebox{0.9\textwidth}{!}{\begin{minipage}{\textwidth}
\begin{tabular}{|l|l|l|l|l|l|l|l|l|l|}
  \hline
  & \multicolumn{3}{c|}{Bayes Graphon} & \multicolumn{3}{c|}{ANCOVA}& \multicolumn{3}{c|}{Bayes ANCOVA}\\\hline
 Total&&&&&&&&&\\ 
sample size& 2000 & 1000 &500& 2000 & 1000 &500& 2000 & 1000 &500\\\hline
 
 Mean-width &1.33&1.33&1.33&1.64&1.69&1.86&1.63&1.72&1.90\\ \hline Number of&$-644.199$&$-645.43$&$-809.2$&$-260.897$&$-2737.39$&$-3262.96$&$-696.38$&$-711.95$&$-832.11$\\ connections &&&&&&&&&\\
 
   \hline

   \hline
\end{tabular}
\end{minipage}}
\label{tablepred40}
\end{table}

\begin{table}[htbp]
\caption{Comparing prediction performances for total sample size 200 across different models for J=80. Prediction MSE is used for mean-width and Predictive mean log-likelihood is used for the number of connections. Half of the recorded sample size is used to estimate model parameters and the rest half is used to evaluate prediction performance.}
\footnotesize\setlength{\tabcolsep}{1pt}
\centering
\begin{tabular}{|l|l|l|l|}
  \hline
   & \multicolumn{1}{c|}{Bayes Graphon} & \multicolumn{1}{c|}{ANCOVA}&\multicolumn{1}{c|}{Bayes ANCOVA}\\\hline
 Mean-width &1.33&2.44& 2.55\\ \hline Number of&$-2704.31$&$-36791.27$&$-3802.34$\\
 connections &&&\\
 
   \hline

   \hline
\end{tabular}
\label{tablepred80}
\end{table}

\begin{figure}[htbp]
\centering
\includegraphics[width=1\linewidth]{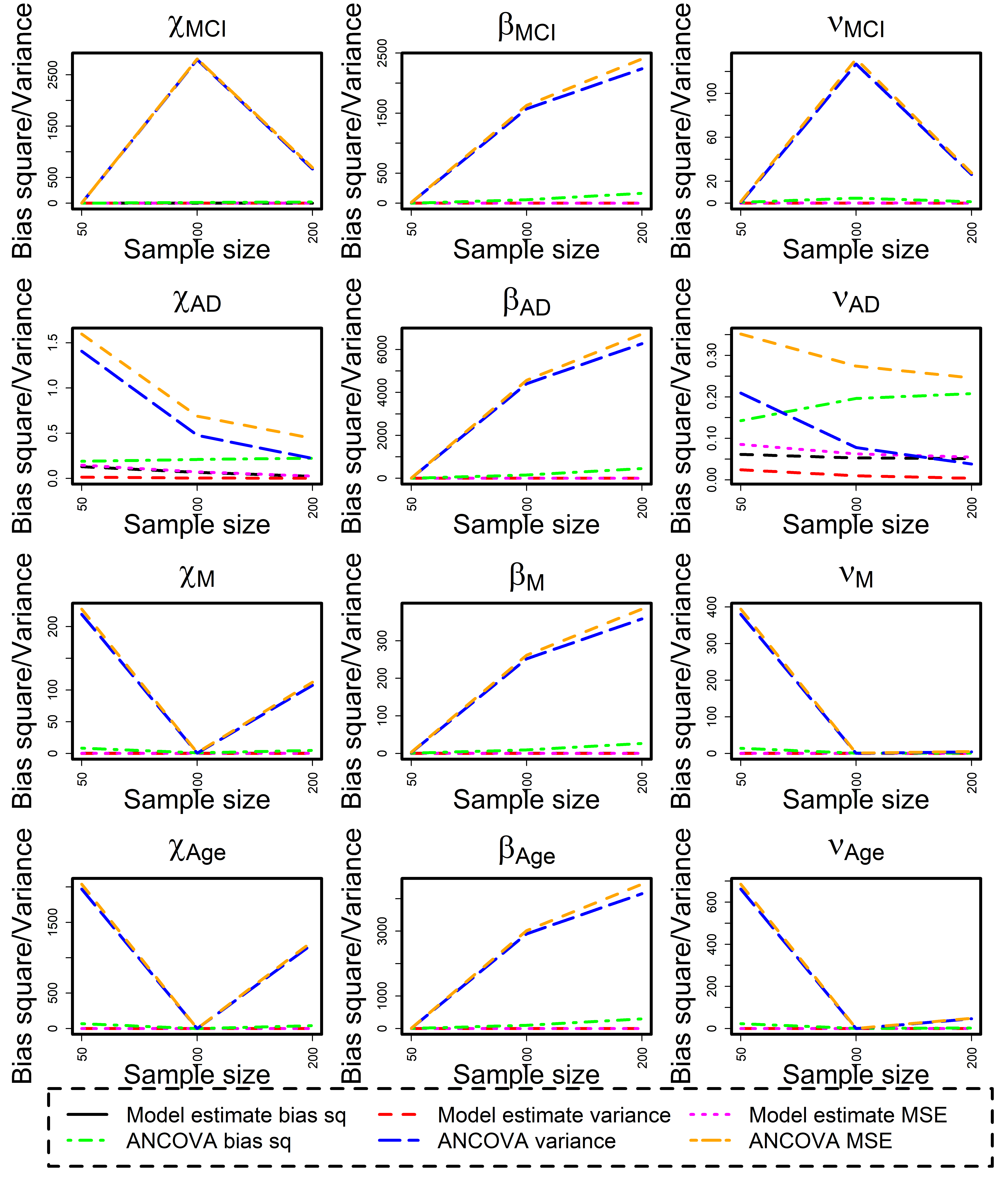}
\caption{Plot of squared bias and variance for different parameters against sample sizes for Bayesian method and ANCOVA estimates for small sample sizes 50, 100 and 200 with $J=20$.}
\label{comparesm}
\end{figure}

\begin{figure}[htbp]
\centering
\includegraphics[width=1\linewidth]{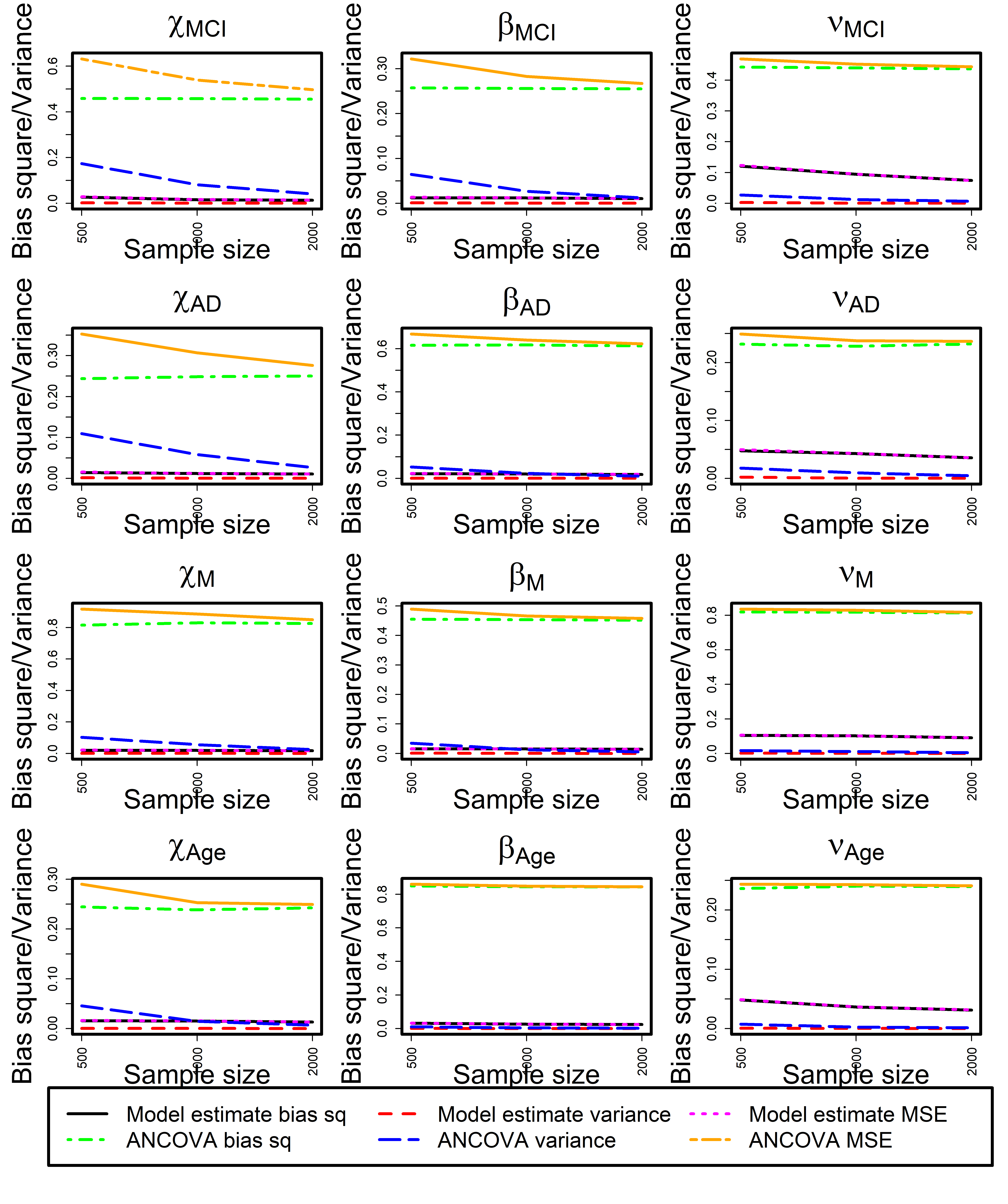}
\caption{Plot of squared bias and variance for different parameters against sample sizes for Bayesian method and ANCOVA estimates for large sample sizes 500, 1000 and 2000 with $J=20$.}
\label{comparela}
\end{figure}

\begin{figure}[htbp]
\centering
\includegraphics[width=1\linewidth]{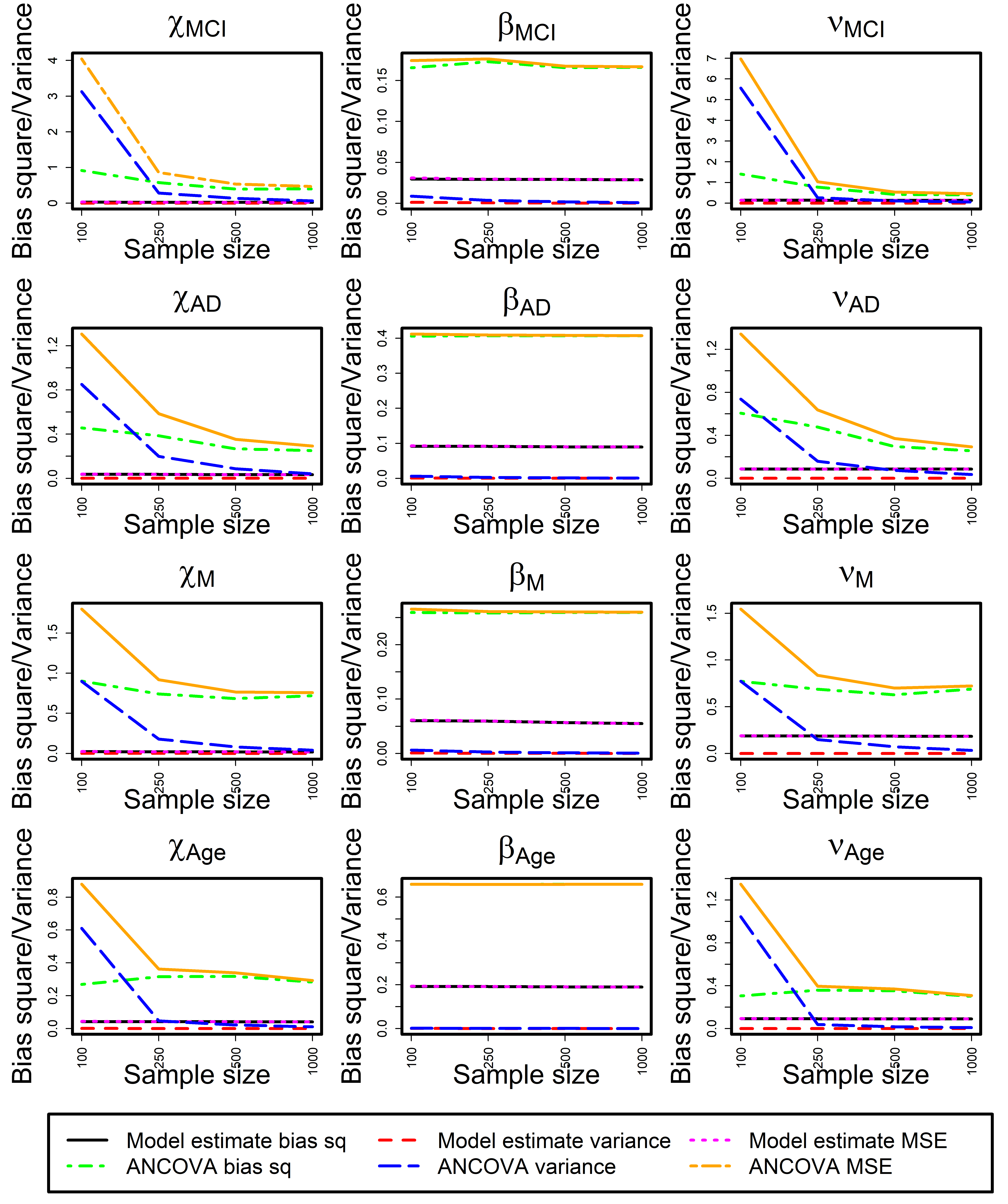}
\caption{Plot of squared bias and variance for different parameters against sample sizes for Bayesian method and ANCOVA estimates for small sample sizes 100, 250, 500 and 1000 with $J=40$.}
\label{compareJ40}
\end{figure}

We present a comparative plot of squared bias, variance, and MSE of the estimates across different sample sizes in Figure~\ref{comparesm} for small sample sizes and Figure~\ref{comparela} for large sample sizes for $J=20$. Fro $J=40$, we present similar plots for sample sizes 100, 250, 500 and 1000 in Figure~\ref{compareJ40}. We have not included the Bayesian ANCOVA estimates in any of these plots as estimation MSEs since the frequentist ANCOVA and the Bayesian ANCOVA estimates are very similar. For the largest sample size 2000 with $J=20$, Table~\ref{tablews} contains the bias--squares and variance values for both of these two methods. These comparisons are also made for the sample size 100 with $J=80$ in Table~\ref{tablews80}. The bias--squares and variances of the estimated matrices are calculated after averaging over bias--squares and variances of the individual entries. In the case of ANCOVA, there are several missing values in the estimates of these matrices. Thus the bias square and variances cannot be calculated at those entries. Thus these are calculated by averaging over only the available ones. In Tables~\ref{tablepred20}, ~\ref{tablepred40} and~\ref{tablepred80}, we present the prediction MSE and the mean predictive likelihood for the logarithm of the connection width and the number of connections, respectively.

In Figure~\ref{heatmapws}, we see that the proposed Bayesian method identifies the true structure in most of the cases. For some cases, it captures the structure but the color levels are different. Usually, the differences are very small as can be observed in Table~\ref{tablews} and~\ref{tablews80}. But for ANCOVA, even for a sample size as large as 2000, it could not capture the true structures. Figure~\ref{comparesm} suggests that bias squares and variances of ANCOVA estimates are not decreasing as sample size increases for small sizes due to the varying missing entries for different sample sizes. As the sample size increases the number of missing values in the estimate goes down. Thus more parameters can be estimated. More parameters become estimable as sample size increases, but this change is not huge. Thus for larger sample sizes, bias--squares and variances of the ANCOVA estimates are decreasing. From Figure~\ref{comparela}, we can also conclude that the proposed Bayesian method performs much better than ANCOVA for sample sizes as large as 500, 1000 or 2000. This is evident from the Table~\ref{tablews} where bias--squares values for the Bayesian estimates are around 40 times smaller and variances are 300 times smaller for the Bayesian estimates than ANCOVA. It may be recalled that estimation by the Bayesian method is consistent. For $J=40$, in Figure~\ref{compareJ40}, we can see that the Bayesian method overwhelmingly outperforms ANCOVA. In both Table~\ref{tablepred20} and Table~\ref{tablepred40}, in terms of prediction accuracy the proposed Bayesian method performs much better than other alternatives.

\section{real-data analysis}
\label{real-data}

\begin{figure}[htbp]
\centering
\includegraphics[width=0.8\linewidth]{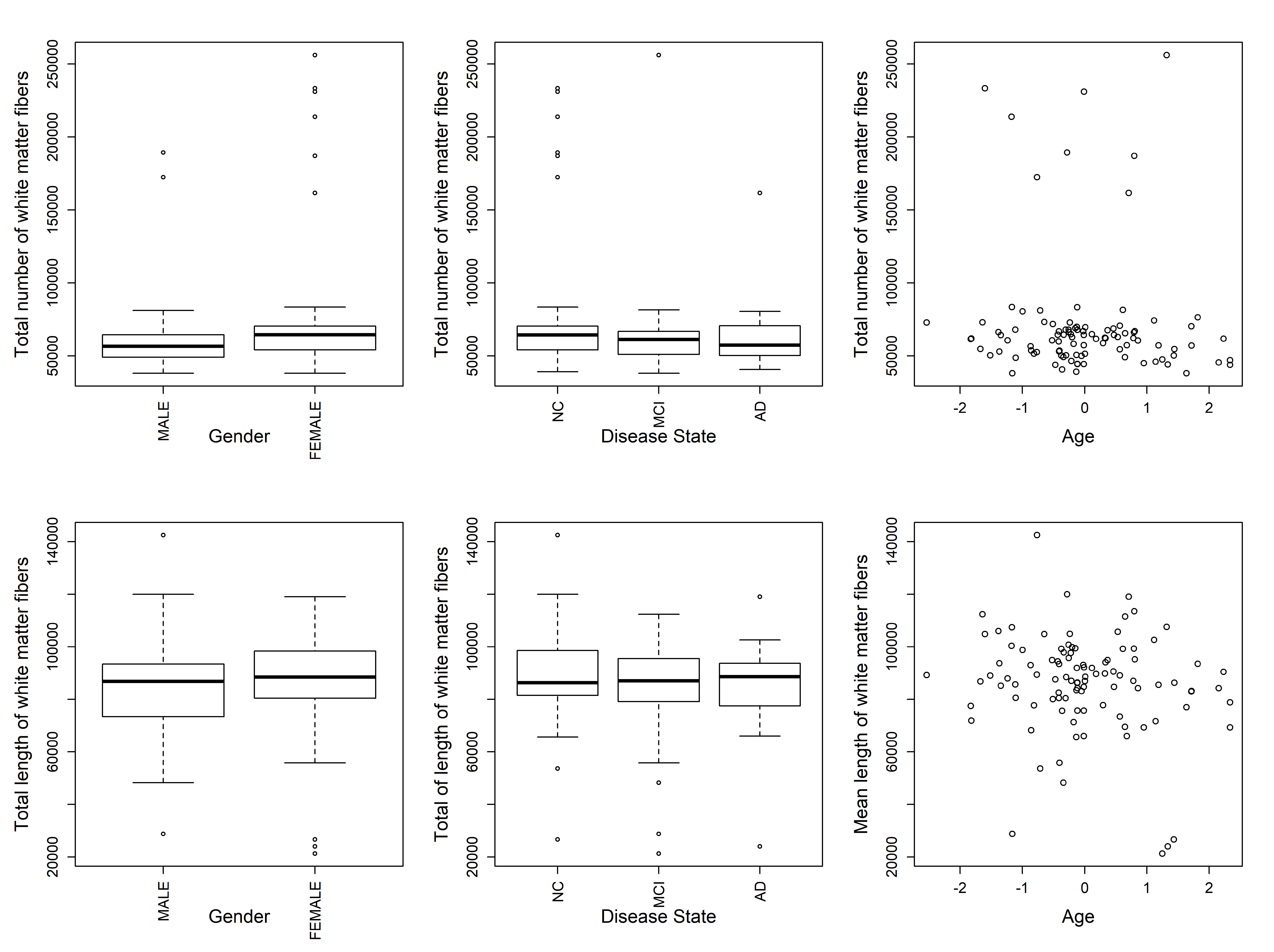}
\caption{Comparison of the distribution of total number of connected white matter fibers and their total length across gender, disease states and age.}
\label{realdataplot}
\end{figure}

We analyze a real-dataset of 100 individuals, collected from ADNI. A demographic summary of the data is provided in Table~\ref{tabledemo}. In disease categories, we have fewer females than males. The baseline subject is a female subject of the average age of 73.9 with no cognitive impairment. In Figure~\ref{realdataplot}, we show that the total number of connected edges and the total length of the fibers vary with gender, disease state, and age. We can see that these distributions are different for male and female. Also, the individuals with Alzheimer's disease seem to have more short-range connections.

\begin{table}[htbp]
\caption{Demographic table}
\centering
\begin{tabular}{rrr}
  \hline
 Covariates& Female & Male \\ 
  \hline
No cognitive impairment (NC) & 17 & 22 \\ 
 Alzheimer's disease (AD) & 6 & 15 \\ 
  Mild cognitive impairment (MCI) & 14 & 26 \\ 
  Average Age & 71.80 & 75.15 \\ 
           & (6.74) & (7.04) \\ 
   \hline
\end{tabular}
\label{tabledemo}
\end{table}

\subsection{Modeling} 

As described in the Section~\ref{prior}, we first generate 10 sets of latent variables for $\xi$ and $\delta$. Then we model the graphon functions with 7 to 20 B-spline basis functions and calculate AIC values using standard R packages for linear and generalized linear regressions, given each set of latent variables. After that, we take the average of the AIC values for each case of a number of B-spline basis functions. After comparing these average AIC values, 13 B-spline basis functions are considered for the graphon functions in estimation for the real-data. Other hyperparameters are kept the same as the simulation i.e. $a= 10$, $M=10$, $b_1=b_2=0.1$ and $c_1=c_2=10$. We collect in total 10000 MCMC samples. Out of those, 5000 are post-burn samples, collected after burning--in the first 5000 samples. We perform a test of significance for each edge by checking if zero is included in the 95\% credible region, constructed from the post-burn samples. After that, we rank those by calculating the probability of greater than zero or less than zero depending on whether zero is in the left tail or in the right tail of the empirical distribution, constructed from the post-burn samples. This can be used as an alternative for the p-value in the frequentist setup. If this comparison is inconclusive, that is we get zero as the probability, we then compare the lengths of the credible sets. Shorter the length, more significant is that edge. This is because a shorter length would suggest more concentration of the posterior distribution around the posterior mean. All of the estimated effects of the covariates for the top ten significant edges are negative for the connection probability i.e. the probit regression part of the model as in Table 5 to 8 in the supplementary material. For the number of connection, the estimated effects corresponding to AD, gender, and age are mostly negative in their corresponding top ten edges. Some estimated effects corresponding to MCI are also negative. These results can be found in Table 9 to 12 in the supplementary materials. For edge-length, all the estimated effects corresponding to gender and age are negative among the top ten significant edges in Table 3 to 4 in the supplementary materials. Some estimated effects of MCI and AD are negative in their corresponding top ten significant edges as in the first two tables of the supplementary materials.

\begin{landscape}
\begin{figure}[htbp]
\centering
\includegraphics[width=1\linewidth]{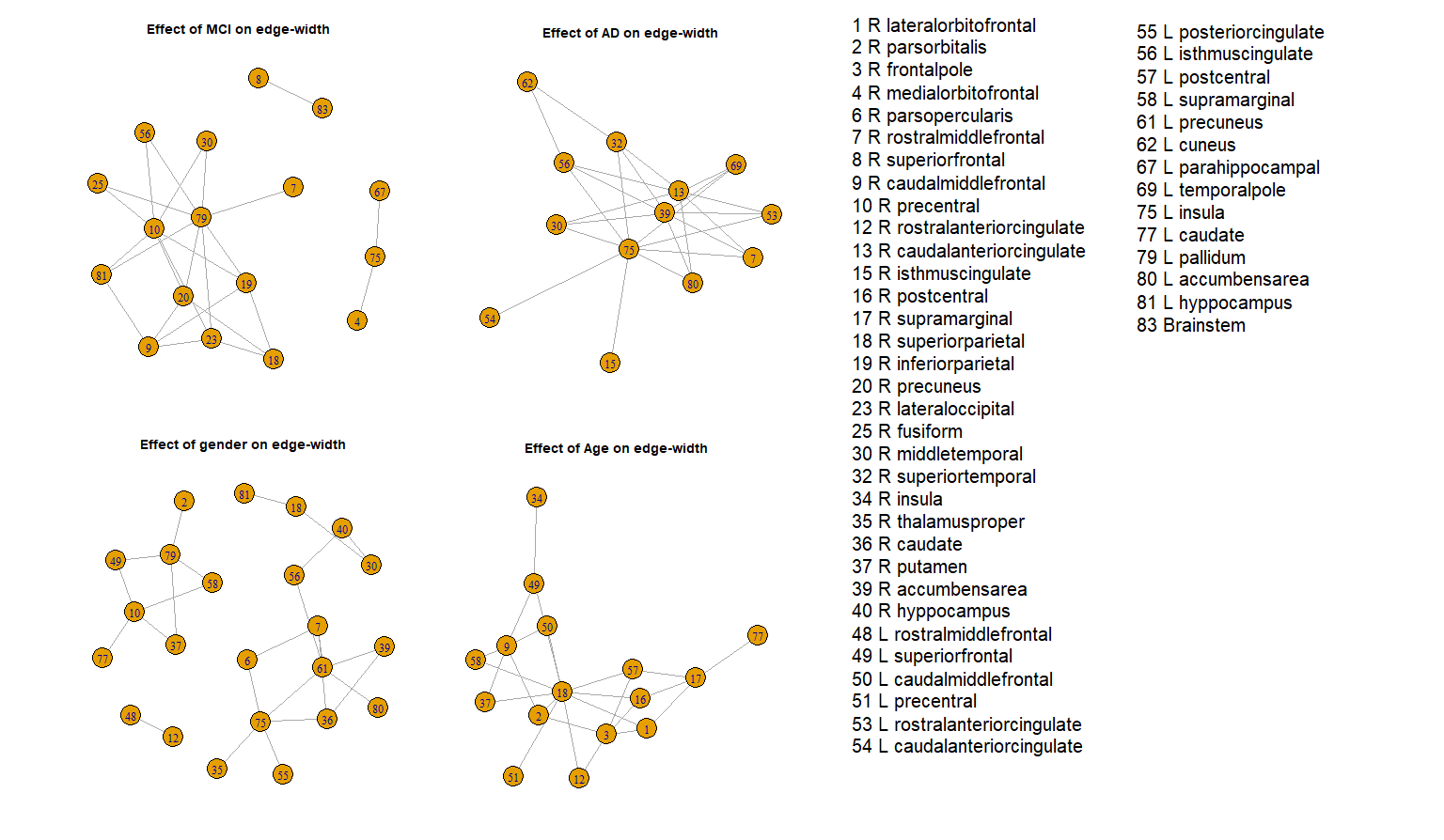}
\caption{Effect of different covariates on average connection length, each circle denotes different cortical brain regions.}
\label{effect of edge-length}
\end{figure}
\end{landscape}

\begin{landscape}
\begin{figure}[htbp]
\centering
\includegraphics[width=1\linewidth]{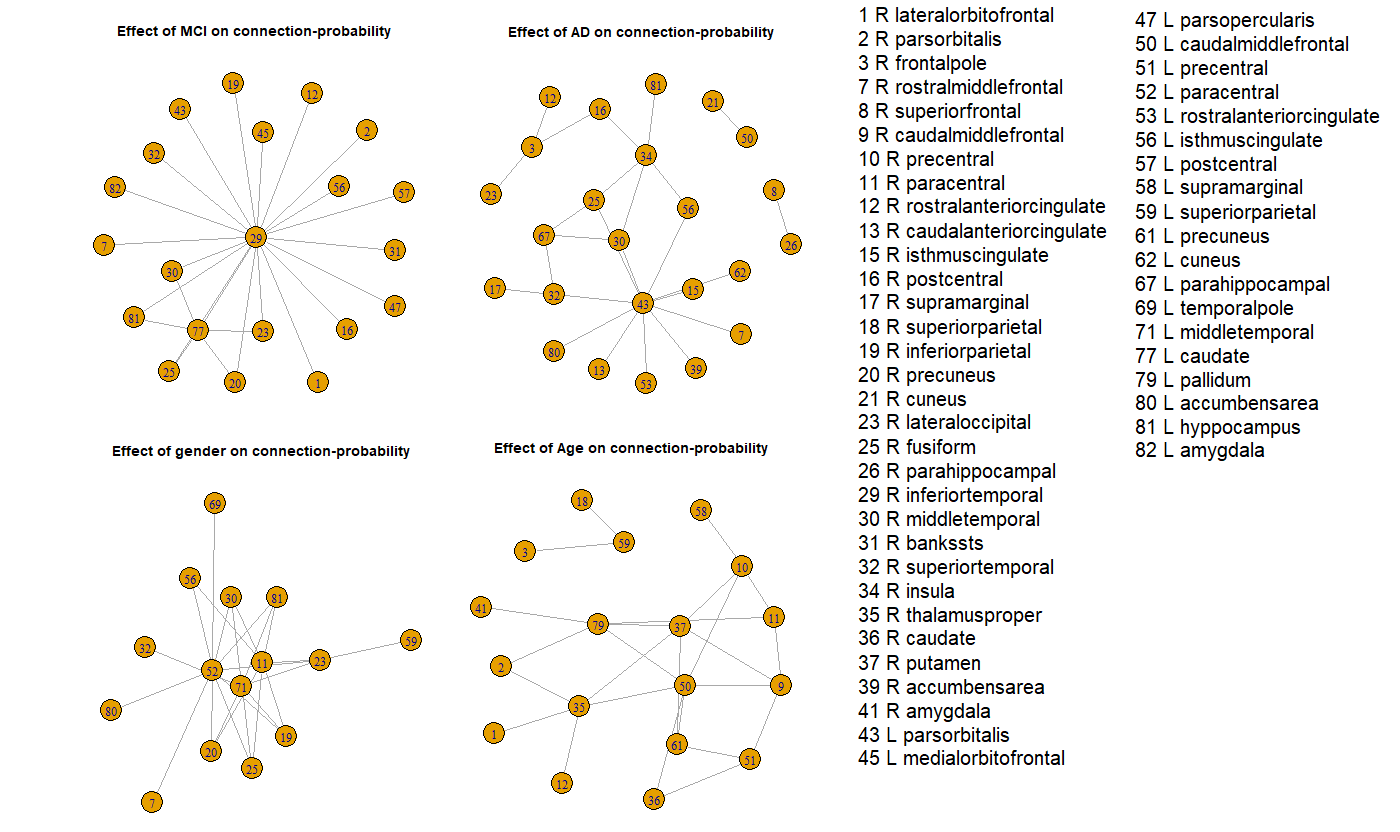}
\caption{Effect of covariates on connection-probability, each circle denotes different cortical brain regions.}
\label{effect of prob-connect}
\end{figure}
\end{landscape}

\begin{landscape}
\begin{figure}[htbp]
\centering
\includegraphics[width=1\linewidth]{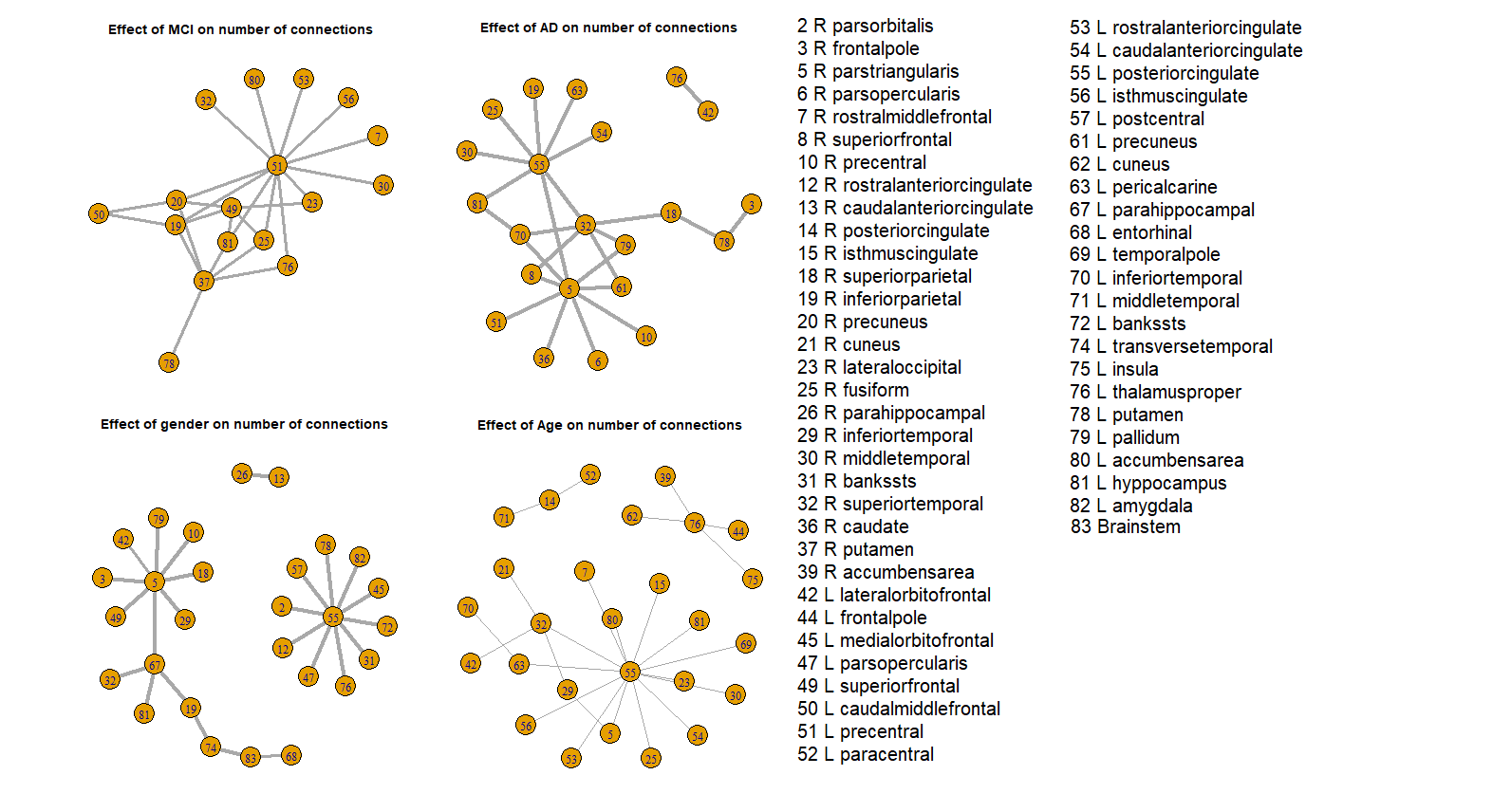}
\caption{Effect of covariates on number of connections, each circle denotes different cortical brain regions.}
\label{effect of connect-num}
\end{figure}
\end{landscape}

Significant edges are plotted for each parameter of interest. 
The circles are the different regions and their names are mentioned in the legend. These plots are in Figures~\ref{effect of edge-length}, \ref{effect of prob-connect} and \ref{effect of connect-num}. We find that for each part of the model, i.e. mean connection length, a number of connections and the binary variable signifying the presence or absence of connection, there are separate sets of regions that have the most number of significant edges, connected with them. In the supplementary materials part, there are additional tables, containing top 10 significant edges for each covariate.

There are some regions that have many connected edges in these plots like insula, pallidum, inferior temporal, parsorbital, precentral, posterior cingulate, superior temporal, superior parietal, middle temporal, paracentral, caudal middle frontal etc. In some of the previous studies on Alzheimer's disease, they were mentioned. Some of those are mentioned in Section~\ref{conclusions-discussion}.

\section{Conclusions and discussion}
\label{conclusions-discussion}

We study the effects of some common measurable covariates on the human brain connectome. Our work extends statistical inference on graph structure from the two sample problem (\cite{Tang})  to a more general regression modeling framework. We propose regression models to explain the extent of connections between different cortical brain regions. In this setup, traditional techniques of separately regressing connections, between each pair of regions and the covariates are not appropriate due to missingness at several edges of the connectome. We solve this problem by using graphon functions to reduce the dimension of the parameter space through a fewer number of fundamental parameters and develop a Bayesian method to estimate those. Subject inhomogeneity is incorporated through random effects and the distributions of the random effects are estimated by using Dirichlet process scale mixture of normal (DPSMN) prior. Figure~\ref{heatmapws} suggests that our Bayesian method identifies the true structure for all the cases in the simulation setting. ANCOVA could not capture the true structures even when the sample size is as large as 2000. Figure~\ref{comparesm} suggests that the small sample performance of ANCOVA is poor. The variance of the ANCOVA estimate is the key issue here. Even for large sample sizes, we conclude that the proposed Bayesian method performs better than ANCOVA as shown in Figure~\ref{comparela}.

The regions that have many connected edges are summarized here. We also cite next to each region's name previous literature that studied those regions in the context of Alzheimer's disease or dementia and found significant. These regions are insula \citep{bonthius2005pathology}, precuneus \citep{karas2007precuneus,klaassens2017diminished}, pallidum \citep{lehericy1991cholinergic}, inferior temporal \citep{scheff2011synaptic}, parsorbital \citep{mclimans2016novel}, precentral \citep{canu2011mapping}, posteriorcingulate \citep{leech2013role}, superior temporal \citep{gao2018aberrant}, superior parietal \citep{migliaccio2015mapping}, middle temporal \citep{jack1998rate}, paracentral \citep{karavasilis2017specific} and caudal middle frontal \citep{bakkour2013effects}.

The greater magnitude of the negative effect on a number of white matter fiber connections between regions for AD subjects than for MCI subjects suggests that white matter connectivity is progressively degraded by lost connections across the clinical spectrum of dementia. In the current analysis, the most significant relationships between a number of white matter connections and MCI or AD status are among widely distributed regions of the brain. In particular, they include connections between regions in the right hemisphere and left hemisphere. They also predominantly involve the frontal, temporal and parietal lobes, and the cingulate cortex. These regions are generally involved with the widespread damage associated with the AD. The preferential loss of ``long-range" fibers (i.e., fibers between the cerebral hemispheres) in these widely distributed regions corresponds with prior work which has demonstrated that long-range connections become degraded with progression along the AD spectrum, leaving highly connected and short-range ``hub" networks relatively intact until late in the disease course \citep{gao2014relevance,sanz2010loss}. 

\section{Acknowledgments}

Data collection and sharing for this project was funded by the Alzheimer's Disease Neuroimaging Initiative (ADNI) (National Institutes of Health Grant U01 AG024904) and DOD ADNI (Department of Defense award number W81XWH-12-2-0012). ADNI is funded by the National Institute on Aging, the National Institute of Biomedical Imaging and Bioengineering, and through generous contributions from the following: AbbVie, Alzheimer's Association; Alzheimer's Drug Discovery Foundation; Araclon Biotech; BioClinica, Inc.; Biogen; Bristol-Myers Squibb Company; CereSpir, Inc.; Cogstate; Eisai Inc.; Elan Pharmaceuticals, Inc.; Eli Lilly and Company; EuroImmun; F. Hoffmann-La Roche Ltd and its affiliated company Genentech, Inc.; Fujirebio; GE
Healthcare; IXICO Ltd.; Janssen Alzheimer Immunotherapy Research \& Development, LLC.; Johnson \& Johnson Pharmaceutical Research \& Development LLC.; Lumosity; Lundbeck; Merck \& Co., Inc.; Meso
Scale Diagnostics, LLC.; NeuroRx Research; Neurotrack Technologies; Novartis Pharmaceuticals Corporation; Pfizer Inc.; Piramal Imaging; Servier; Takeda Pharmaceutical Company; and Transition
Therapeutics. The Canadian Institutes of Health Research is providing funds to support ADNI clinical sites in Canada. Private sector contributions are facilitated by the Foundation for the National Institutes of Health (\url{www.fnih.org}). The grantee organization is the Northern California Institute for Research and Education,
and the study is coordinated by the Alzheimer’s Therapeutic Research Institute at the University of Southern California. ADNI data are disseminated by the Laboratory for Neuro Imaging at the University of Southern California.

We are grateful to two anonymous reviewers, the associate editor and the editor for their valuable comments that have greatly helped to improve the manuscript.

\bibliographystyle{imsart-nameyear}
\bibliography{main}

\end{document}